\documentclass[12pt,preprint]{aastex}
\usepackage{graphicx}

\slugcomment{Astrophysical Journal Letters {\it in press}}
\shorttitle{Measurement of Photospheric Neon}
\shortauthors{J.J.~Drake \& B.~Ercolano}
\begin{document}

\title{The Detectability of Neon Fluorescence and Measurement of the
  Solar Photospheric Neon Abundance} 

\author{Jeremy J.~Drake\altaffilmark{1}}
\author{Barbara Ercolano\altaffilmark{1}}
\affil{$^1$Smithsonian Astrophysical Observatory,
MS-3, \\ 60 Garden Street, \\ Cambridge, MA 02138}
\email{jdrake@cfa.harvard.edu}

\begin{abstract}
Monte Carlo calculations of the Ne~K$\alpha$ line fluoresced by
coronal x-rays and emitted near the temperature minimum region of the
solar atmosphere have been employed to investigate the use of this
feature to measure directly the solar photospheric Ne abundance.
Though very weak, comparison with spectral line databases indicates
that at plasma temperatures typical of the quiet Sun and cool active
regions ($\leq 2\times 10^6$~K) the line is isolated and unblended.  A
canonical solar chemical composition yields an equivalent width of
$\sim 6$~m\AA\ (0.3~eV) when observed at heliocentric angles $\sim 0$.
For a 1~arcmin field of view, photon fluxes at Earth are of order
0.2~ph~s$^{-1}$ for the quiet Sun, rendering the Ne~K$\alpha$
fluorescent line a quite feasible means for determining the solar
photospheric Ne content.
\end{abstract}


\section{Introduction}
\label{s:intro}

The abundance of the element neon is notoriously difficult to
determine in cool stars like the Sun.  Ne exhibits no lines in the
visible light spectra of late-type stars, and the solar abundance is
based largely on transition region and coronal lines and energetic
particle measurements, supplemented with local cosmic estimates
\citep[e.g.\ ][]{Meyer:85b,Anders.Grevesse:89}.  The Ne abundance is
usually measured relative to that of O \citep[see, however, the recent
 Ne/H measurement of][]{Landi.etal:07}.  While the majority of past
solar Ne/O estimates \citep[see the supplementary data
in][]{Drake.Testa:05}, together with recent analyses of Solar Maximum
Mission and Solar and Heliospheric Observatory archival spectra
\citep{Schmelz.etal:05,Young:05b}, support the canonical
ratio of 0.15 
by number \citep[e.g.][]{Anders.Grevesse:89,Asplund.etal:05}, there is
a scatter of measurements around this value by more than a factor of
2---even among measurements of different solar regions made with the
same instrumentation \citep{McKenzie.Feldman:92,Strong.etal:88}.
Based on an analysis of X-ray He-like Ne and H-like O resonance lines
\citet{McKenzie.Feldman:92} concluded that the Ne abundance must vary
in the solar corona.
%, a conclusion reinforced by the re-analysis by \citet{Drake:07} of
%the same data, together with fluxes for the same transitions obtained
%from {\em Solar Maximum Mission} spectra by \citet{Schmelz.etal:06}.
Faced with the prospect of coronal Ne fractionation by a process that
is not yet firmly identified or understood, it is not clear that the
neon content of {\em any} region of the solar outer atmosphere will be
the same as that of the deeper layers.

The solar neon content represents a potentially large source of
uncertainty for understanding the oscillation spectrum of the Sun.
Models employing a recently advanced solar chemical composition based
on 3-D non-LTE hydrodynamic photospheric modelling
\citep{Asplund.etal:05} lead to predictions of the depth of the
convection zone, helium abundance, density and sound speed in serious
disagreement with helioseismology measurements
\citep{Basu.Antia:04,Bahcall.etal:05}.  The \citet{Asplund.etal:05}
mixture contains less of the elements C, N, O and Ne that are
important for the opacity of the solar interior by 25-35~\%\ compared
to earlier assessments
\citep[e.g.][]{Anders.Grevesse:89,Grevesse.Sauval:98}.
\citet{Antia.Basu:05,Bahcall.etal:05c} suggested the uncertain solar
Ne abundance might be raised to compensate.  While enthusiasm for this
solution has been dampened by a study of solar parameter uncertainties
inferred from oscillation data that appears to exclude such large Ne
abundance revisions \citep{Delahaye.Pinsonneault:06},
\citet{Drake.Testa:05} found empirical support from {\it Chandra} high
resolution X-ray spectra of mostly magnetically active stars for which
the Ne/O abundance ratio appears consistently higher by a factor of
$\sim 2$ or more than the currently recommended solar value of
Ne/O=0.15 by number.

In this {\em Letter} we examine the possibility of using the
Ne~K$\alpha$ X-ray line of neutral Ne at 14.61~\AA\ (0.849~keV) to
measure the {\em photospheric} Ne abundance.  \citet{Phillips.etal:94}
used the analogous K$\beta$ transition of Fe to probe 
differences between photospheric and coronal Fe abundances.  While undoubtedly very
weak compared with the lines of highly ionised Ne formed in the
transition region and corona, the Ne~K$\alpha$ line is formed by
inner-shell photoionisation by coronal X-rays of Ne near the
temperature minimum region of the solar atmosphere
\citep{Drake.Ercolano:07}, beneath the chromospheric zone where
chemical fractionation processes related to element first ionisation
potential (FIP) are thought to occur
\citep[e.g.]{Meyer:85b,Feldman:92}.  We present Monte Carlo
calculations of the fluorescent Ne~K$\alpha$ line in \S2, and discuss
its observability in \S3 and \S4.

\section{Monte Carlo Calculations of Neon Fluorescence}
\label{s:monte}

The X-ray Ne ``characteristic'' (fluorescence) K$\alpha$ line at
14.61~\AA\ \citep{Bearden:67} corresponds to the $2p-1s$ decay of the
excited state resulting from ejection of an inner-shell $1s$ electron
in neutral or near-neutral Neon by either electron impact or
photoionisation.  In the case of the solar photosphere illuminated
from above by coronal X-rays, fluorescent lines will be produced
almost entirely by photoionisation
\citep{Basko:78,Bai:79,Parmar.etal:84}.  \citet{Bai:79} pointed out
that, for a given source spectrum, $F(\lambda)$, the observed flux of
K$\alpha$ photons from the photosphere depends on essentially three
parameters: the photospheric abundance $A$ of the fluorescing species
relative to that of other elements of significance for the
photoabsorption opacity in the vicinity of the $1s$ ionisation edge;
the height $h$ of the emitting source; and the heliocentric angle
$\theta$ between the emitting source and the observer.

Fluorescent lines are formed in the region of an atmosphere
corresponding to optical depth unity for the primary K-shell ionising
photons.  \citet{Drake.Ercolano:07} showed that in the case of the
fluorescent lines from abundant elements O-Fe formed in the solar
atmosphere, this occurs below the chromosphere.  For the case of Ne
the solar atmospheric Model C (VALC) of \citet*{Vernazza.etal:81}
indicates that the K-shell $\tau=1$ depth occurs at a gas temperature
of about 5000~K just above the temperature minimum and about 700~km
above the point where the continuum optical depth at 5000~\AA,
$\tau_{5000}$, is unity.  
%The observed Ne line strength is primarily
%sensitive to the ratio of the Ne~K photoabsorption cross-section
%relative to that of all other elements at the depth of
%formation---primarily O.

To estimate the expected intensity of the emergent Ne~K$\alpha$ line
we used a modified version of the 3D Monte Carlo radiative transfer
code MOCASSIN \citep{Ercolano.etal:03,Ercolano.etal:05}.  This code
has been tested in detail for Fe~K$_\alpha$ photospheric fluorescence
problems by comparison with the computations of \citet[][see
  \citealt{Drake.Ercolano:07}]{Bai:79}.  Computation of Ne~K
fluorescence is similar to that for Fe~K and we describe our method
here only in brief; the reader is referred to the earlier work for
further details.

The fluorescence calculation involves following the fate of
monochromatic energy packets that sample the spectrum of the overlying
corona and that are incident on the photosphere.  We assume the
photosphere to be ``cold'', whereby all elements, including Ne, are
neutral.  Energy packets can undergo photoabsorption or Compton
scattering, the probabilities of which are determined by the
respective cross-sections.  Photoabsorption by Ne of a packet with
energy above the K-shell ionisation threshold (870~eV) is
immediately followed by re-emission of $n$ of Ne\,K${\alpha}$ packets
from the same event location.  For an incident packet of frequency
$\nu$, carrying energy $\varepsilon_0$ in the unit time $\Delta\,t$
the total Ne\,K${\alpha}$ emission is given by
\begin{equation}
L(Ne\,K{\alpha})~=~n
L'(Ne\,K{\alpha})~=~n\,\frac{\varepsilon_{0}}{\Delta\,t}
\frac{1}{h\nu}\,
\frac{\kappa_{\nu}^{Ne}Y_{Ne\,K{\alpha}}\varepsilon_{Ne\,K{\alpha}}}
{\kappa_{\nu}^{gas}}R_{\alpha}
\end{equation}
where $\kappa_{\nu}^{Ne}$ and $Y_{Ne\,K{\alpha}}$ are the
absorption opacity and the Ne\,K${\alpha}$ yield,
$\varepsilon_{Ne\,K{\alpha}}$ is the energy of the K$\alpha$ line 
($\sim$0.848\,keV), $\kappa_{\nu}^{gas}$ is the absorption
opacity due to all other abundant species and $R_{\alpha}$ is the
branching ratio between $K{\alpha}$ and $K{\beta}$ fluorescence
(0.882:0.118, \citealt{Bambynek.etal:72}).  We adopted a value of
0.018 for the fluorescence yield of neutral neon \citep{Krause:79}.

The fates of Ne\,K${\alpha}$ packets are then determined by the absorption
and Compton opacities encountered along their diffusion paths.  
Emergent integrated and direction-dependent spectral energy
distributions are determined from the packets that escape the
photosphere.  For a
given coronal X-ray spectrum, the maximum intensity of a fluorescent
line is achieved for a heliocentric angle $\theta=0$ and coronal
height $h=0$.  Since we are primarily interested here in whether the
line is observable or not, we adopt these as baseline parameters in
order to estimate the maximum possible strength of the line.

The Ne~K$\alpha$ flux was computed for isothermal irradiating coronal
spectra with plasma temperatures in the range $10^6$-$10^7$~K and the
chemical composition of \citet[][with Ne/H=8.08 on the usual log+12
scale]{Grevesse.Sauval:98}.  We also performed calculations for a Ne
abundance elevated by a factor of 3 (Ne/H=8.58, or [Ne/H]=0.5).  The
calculated Ne~K$\alpha$ flux is sensitive to the chemical mixture in
the photosphere, but is also sensitive to some extent to that assumed
for the coronal spectrum through the contribution to the ionising flux
from lines (see \S\ref{s:mcresults}).  Coronal spectra were computed
using emissivities from the CHIANTI database (v5.2;
\citealt{Dere.etal:97,Landi.etal:06}) and the ion populations
of \citet{Mazzotta.etal:98}, as implemented in the
PINTofALE\footnote{The Package for INTeractive Analysis of Line
Emission is freely available from
http://hea-www.harvard.edu/PINTofALE/} IDL suite of programs
\citep{Kashyap.Drake:00}.

\section{Strength of the Ne K$\alpha$ Line}
\label{s:nekstrength}

\subsection{Monte Carlo Results}
\label{s:mcresults}

\begin{figure}
\begin{center}
\includegraphics[width=0.80\textwidth]{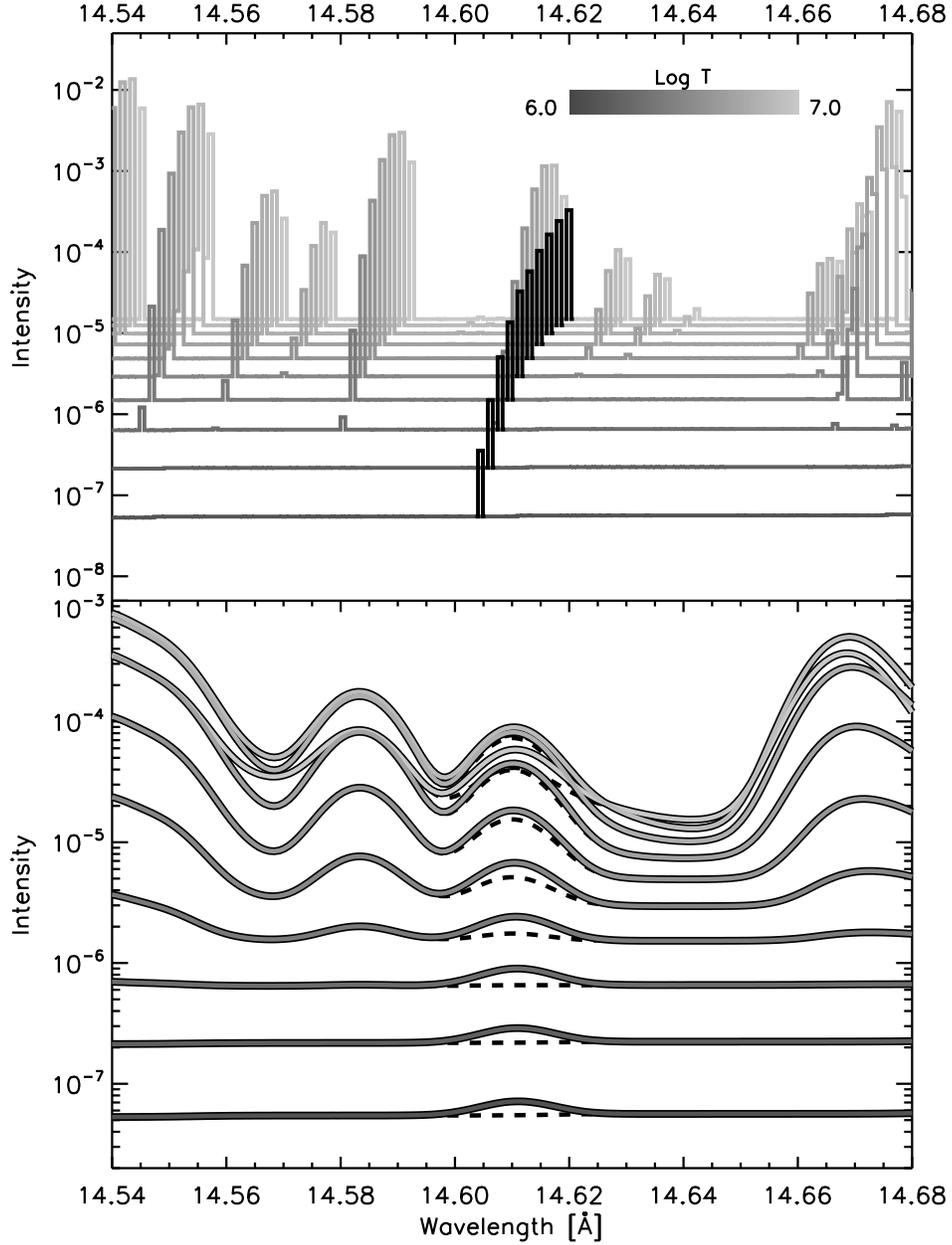}
\end{center}
\caption{Top: The strength of the Ne~K$\alpha$ line (black) 
shown in comparison to
neighbouring and blending lines (grey) in the fluorescing coronal X-ray
spectrum in units of $10^{-23}$~erg~cm$^3$~s$^{-1}$~bin$^{-1}$ (with
bins of size $8.6\times 10^{-4}$~\AA)
for coronal isothermal plasma temperatures in the range
$10^6$--$10^7$~K.  Bottom: The same coronal spectra with (solid
curves) and without (dashed) the addition of Ne~K$\alpha$ smoothed to
a resolving power (FWHM) of $\lambda/\Delta\lambda=1000$.
}
\label{f:spectra}
\end{figure}

Model spectra corresponding to the combination of coronal direct and
photospheric reprocessed X-rays seen by an observer at $\theta=0$ in
the vicinity of the Ne~K$\alpha$ line are illustrated for a range of
coronal temperatures in Figure~\ref{f:spectra}.  Also shown are the
same spectra seen at a resolving power of
$\lambda/\Delta\lambda=1000$, where $\Delta\lambda$ is assumed to be
the full-width at half-maximum of a Gaussian instrument response
function.

From Figure~\ref{f:spectra} it is apparent that the Ne~K$\alpha$ line
coincides rather closely with a 
weak line of Fe~XVIII at 14.61~\AA.  While absent at quiescent plasma
coronal temperatures found in the Sun of 1-$2\times 10^6$~K, this 
line becomes problematic for temperatures significantly above this
range, such as might be found in active regions or
flaring conditions.  The full list of lines in the CHIANTI v5.2 database
within $5\sigma$ ($\pm 0.031$~\AA) of
Ne~K$\alpha$ are listed in Table~\ref{t:5sigma}.  The NIST Atomic
Spectra Database (version 3.1.2; \citealt{Ralchenko.etal:07}) also
lists eight other transitions within the $5\sigma$ range from highly
ionised iron-group elements Ti, V, Cr, and Co.  The brighter lines of
Ti and Cr are included in the CHIANTI database and owing to the low
cosmic abundance of these four elements (300, 3000, 70 and 380 times
less abundant than Fe, respectively; \citealt{Grevesse.Sauval:98}) we
do not anticipate any of these transitions to be of significant
strength compared with the Fe lines or Ne~K$\alpha$.

\begin{figure}
\begin{center}
\includegraphics[width=0.80\textwidth]{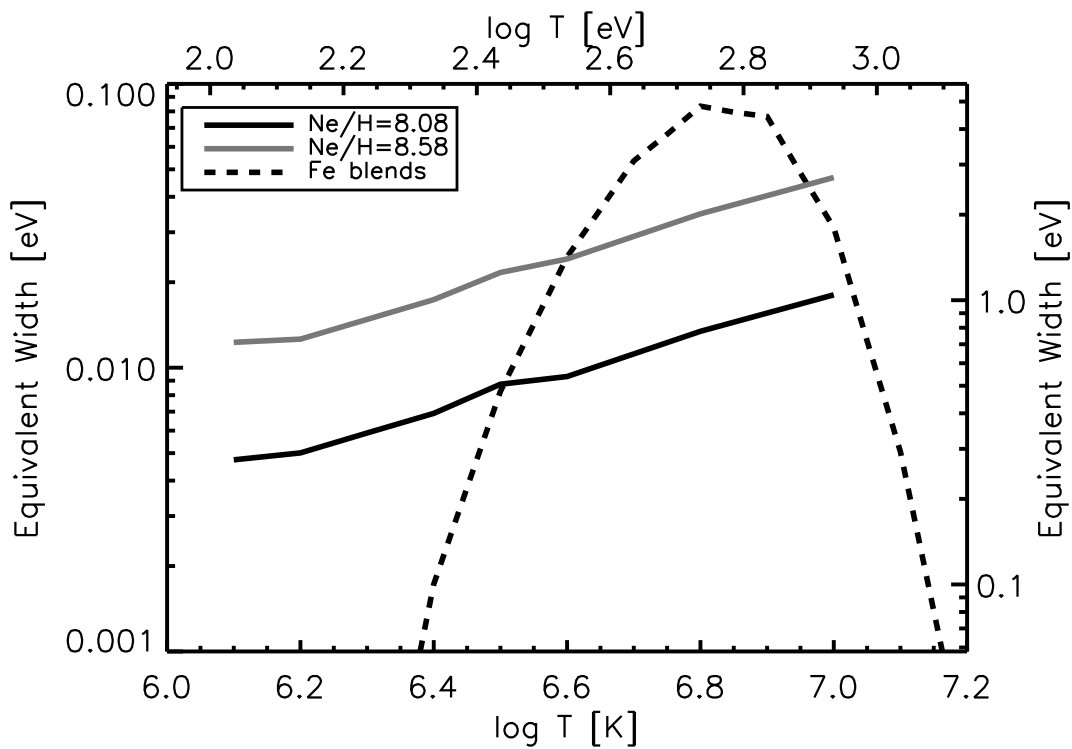}
\end{center}
\caption{The equivalent width of the Ne~K$\alpha$ line with respect to
  the ionising coronal X-ray spectrum as a function of isothermal
  plasma temperature, computed for 1 and 3 times the photospheric Ne
  abundance of \citet{Grevesse.Sauval:98}.    Also shown is the 
  equivalent width of the Fe~XVIII line that blends
  with the Ne feature.  The different y-axes are
  labelled in \AA\ and eV units.}
\label{f:ew}
\end{figure}

A fiducial for the observability of a spectral line can be expressed
in terms of its equivalent width.  In the case of lines excited by
fluorescence, this equivalent width is most usefully related to the
ionising coronal spectrum against which it will be observed, as
illustrated in Figure~\ref{f:spectra}.  
We have calculated this quantity
for each of the models in our grid, for Ne abundances equivalent to
the \citet{Grevesse.Sauval:98} value and for three times this ([Ne/H]=0 and
[Ne/H]=0.5); results are illustrated in Figure~\ref{f:ew}.  Also shown
is the equivalent width for the Fe~XVIII blend.
Since the Ne and Fe~XVIII lines are coincident, the
resolving power required to separate them can be considered both out
of reach of foreseeable instrumentation and physically infeasible
owing to thermal broadening that will irretrievably smear the lines
together.  Nevertheless, at cooler
coronal temperatures the blend is negligible and
Ne~K$\alpha$ should be the only significant spectral feature in the
vicinity.

We conclude that Ne~K$\alpha$ is in principle quite observable and
isolated in high resolution spectra with sufficient signal-to-noise
(S/N) to detect the line above the continuum, provided plasma
temperatures are lower than $\sim 2.5\times 10^6$~K.  If the strength
of the blending Fe~XVIII line can be accurately modelled,
the useful temperature range of Ne~K$\alpha$ measurements could extend
to $\sim 3\times 10^6$~K.

\subsection{Existing Solar Spectra of the Ne~K$\alpha$ Region}

There have been no recent high resolution X-ray spectra taken in the
14-15~\AA\ range of the solar corona.  The most extensive set of
observations dates back to the SMM Flat Crystal Spectrometer
\citep{Acton.etal:80}.  Owing to the limited detector form factors
available, this was a scanning instrument in which only a small
fraction of the spectrum could be recorded at a time, limiting
the exposure and S/N attainable for a given wavelength.  In this
context, it is important to realise that much higher quality spectra
could be obtained with current technology.  Indeed, it is by no means
an exaggeration to note that spectra of this region obtained for much
more distant cosmic X-ray sources by the {\it Chandra} X-ray
Observatory are now routinely of similar or higher spectral quality
than those that have been obtained to date for the non-flaring Sun.

\begin{figure}
\begin{center}
\includegraphics[width=0.80\textwidth]{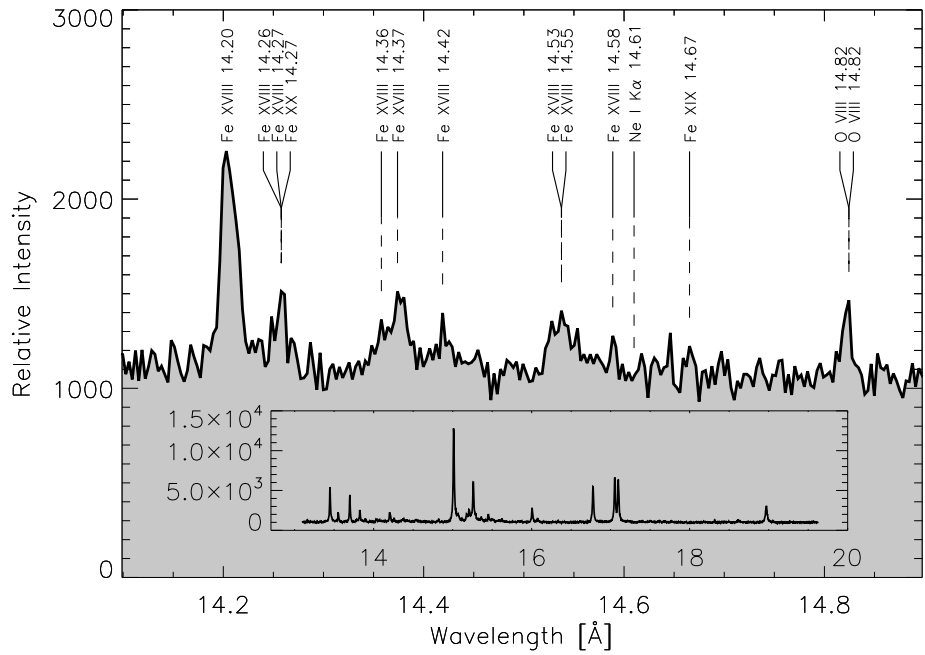}
\end{center}
\caption{The SMM FCS Channel 1 spectrum in the region of the Ne~K$\alpha$ line 
obtained from combining 20
  different observations of cooler active regions.  The full Channel 1
spectrum is shown inset.
}
\label{f:fcs}
\end{figure}

FCS spectra covering our spectral range of interest for cooler solar
active regions have recently been analysed by \citet{Schmelz.etal:05},
who selected 20 of the highest S/N spectra in the $\sim
13$--20~\AA\ range (FCS Channel 1), obtained in 1986-1987, in which
Fe~XIX emission was not visible.  Since these selection criteria are
not too dissimilar to our requirements for seeing unblended
Ne~K$\alpha$, we have examined the same set of spectra obtained from
the FCS archive, summed in order to realise the maximum possible S/N.
The resulting spectrum is illustrated in Figure~\ref{f:fcs}.  An
Fe~XVIII line at 14.20~\AA\ is clearly visible in these spectra: the
intensity of the blending Fe~XVIII line at 14.61~\AA\ can be put in
context by noting that the CHIANTI emissivity ratio for these lines is
52:1.  

%Features seen in Figure~\ref{f:fcs} near 14.61~\AA\ are therefore
%noise and the Fe~XVIII 14.61~\AA\ and any Ne~K$\alpha$ flux is
%essentially invisible at the illustrated scale.

Unfortunately, the FCS spectra suffer from quite severe background
contamination due to fluorescence in the spectrometer crystals
\citep{Phillips.etal:82}.  At the resolution of the FCS channel 1,
corresponding to a resolving power of $\sim 1000$, our synthetic
coronal spectra have peak line-to-continuum flux density ratios of
order $10^3$ for the strong Fe~XVII~15.01~\AA\ line for temperatures of
2--$3\times 10^6$~K.  In the summed FCS spectrum in
Figure~\ref{f:fcs}, the line peaks at $1.3\times 10^4$ in the same
relative flux density units.  The background continuum in this
spectral region is about $1.05\times 10^3$, or $\sim 2$ orders
of magnitude larger than the plasma thermal continuum.  At the same
spectral resolution, a line with intensity corresponding to our
computed Ne~K$\alpha$ equivalent widths---of order 0.35~eV or 6~m\AA\
for a coronal temperature of $2\times 10^6$~K---would have a peak
intensity of order 50\%\ of the thermal continuum intensity: the line
is completely swamped in the FCS by the crystal fluorescence
background.

\subsection{Observability of Ne~K$\alpha$ Fluorescence}

From the above, it is apparent that detection of the Ne~K$\alpha$ line
requires much greater instrumental sensitivity than afforded by the
SMM FCS.  The temperature range in which the line remains blend-free
and isolated is typical of those found in the coolest active regions
and in the quiet Sun.  Quiet Sun X-ray surface fluxes are of order
$10^5$~erg~cm$^2$~s$^{-1}$ \citep[e.g.\ ][]{Withbroe.Noyes:77}.  Our model
coronal X-ray spectra at a temperature of $1.5\times 10^6$~K scaled to
this surface flux and covering, for example, a $1\times 1$~arcmin
field of view corresponds to a flux density in the continuum at
14.61~\AA\ of about 30~ph~s~\AA$^{-1}$, and a Ne~K$\alpha$ flux of
$\sim 0.2$~ph~s$^{-1}$.  High-resolution imaging would not be required
for a useful Ne~K$\alpha$ measurement and such a flux level is in
principle easily observed.  The large intensity contrast between
Ne~K$\alpha$ and neighbouring strong lines, such as Fe XVII~15.01~\AA,
does, however, impose tight instrumental profile requirements in order
to avoid spill-over contamination.  This requirement
could be eased by adopting a narrow-band filter to attenuate
neighbouring strong lines. 

The feasibility of observing Ne~K$\alpha$ also depends critically on
the absence of blending lines.  The veracity of current line lists in
the immediate vicinity of Ne~K$\alpha$ should be investigated further
using appropriate computational and laboratory surveys in order to
evaluate the current assessment that the line is unblended at quiet
Sun coronal temperatures.

%; were spatial discrimination required
%for smaller regions of the solar surface usingwould be easily within
%reach of, for example, a collimator-fed spectrograph or microcalorimeter. 

\section{Conclusion}

We have investigated the observability of the Ne~K$\alpha$ line by
means of 3D Monte Carlo calculations using the MOCASSIN code.
Resulting line fluxes are similar to those of the underlying continuum
of the fluorescing spectrum.  While very faint compared to the
prominent lines of abundant coronal ions, Ne~K$\alpha$ should be
essentially unblended at typical quiet Sun temperatures, but will
become catastrophically blended with a line of Fe~XVIII at
significantly higher temperatures.  We estimate an Ne~K$\alpha$ flux
of $\sim 0.2$~ph~s$^{-1}$ for a 1~arcmin field of the quiet Sun at
heliocentric angles close to $0^\circ$.  Such a flux is easily
observed and the Ne~K$\alpha$ fluorescent line would seem to
represent a feasible means to measure the photospheric Ne abundance.

\acknowledgments

We extend warm thanks to Ken Phillips for help and advice in analysis
of SMM FCS spectra, and to Brian Dennis and Kim Tolbert for assistance
in accessing the FCS archive.  We thank the CHIANTI team for making
available an invaluable compilation of atomic data, and the NASA AISRP
for providing financial assistance for the development of the
PINTofALE package.  JJD was funded by NASA contract NAS8-39073 to the
{\em Chandra X-ray Center} during the course of this research and
thanks the Director, H.~Tananbaum, for continuing support and
encouragement.  BE was supported by {\it Chandra} Grants GO6-7008X and
GO6-7098X.  Finally, we thank the referee, Dr.~Enrico Landi, whose
comments enabled us to improve the manuscript.

%%%%%%%%%%%%%%%%%%%%%%%%%%%%%%%%%%%%%%%%%%%%%%%%%%%%%%

%\bibliographystyle{apj}
%\bibliography{fluor,neon,neon_suppl,jjdrake,chianti,atomic,mocassin}

\newpage

\begin{table}
\caption{Lines in the CHIANTI v5.2 database lying within $5\sigma$ of the
  position of the 14.61~\AA\ Ne~K$\alpha$ transition for a Gaussian FWHM
  resolving power of $\lambda/\Delta\lambda=1000$.
\label{t:5sigma}
}
\begin{tabular}{lcccc}
\hline
Ion & $\lambda$ (\AA) & Rel. Int. & $\log T_{max}$ & Transition \\
\hline
  Fe {\sc XVIII} &  14.584 & 1.0 & 6.90 & 
$2s^22p^5 \; ^2P_{3/2}$ -- $2s^22p^4(^3P)3d \; ^4P_{1/2}$ \\
    Fe {\sc XIX} &  14.596 & $2.5\times 10^{-4}$ & 6.90 & 
$2s2p^5 \; ^1P_{1}$ -- $2s2p^4(^4P)3d \; ^3P_{2}$ \\
    Fe {\sc XIX} &  14.600 & $1.2\times 10^{-4}$ & 6.90 & 
$2s^22p^4 \; ^1S_{0}$ -- $2s^22p^3(^4S)3d \; ^5D_{1}$ \\
  Fe {\sc XVIII} &  14.610  & 0.40 & 6.85 & 
$2s^22p^5 \; ^2P_{1/2}$ -- $2s^22p^4(^3P)3d \; ^2P_{3/2}$ \\
%    Ni {\sc XXI} &  14.616  & 1.6099e-13 & 7.000 & 
%$2s2p^5 \; ^3P_{0}$ -- $2s^22p^3(^4S)3s \; ^3S_{1}$ \\
    Fe {\sc XIX} &  14.622 &  0.03 & 6.90 & 
$2s^22p^4 \; ^3P_{2}$ -- $2s^22p^3(^2D)3s \; ^1D_{2}$ \\
    Fe {\sc XIX} &  14.628 &  0.01 & 6.90 & 
$2s2p^5 \; ^3P_{2}$ -- $2s2p^4(^2D)3s \; ^1D_{2}$ \\
     Fe {\sc XX} &  14.633 & $1.2\times 10^{-3}$ & 7.0 & 
$2s^22p^3 \; ^2P_{3/2}$--  $2s^22p^2(^3P)3s \; ^4P_{1/2}$ \\
\hline
\end{tabular}
\end{table}
\newpage

\end{document}